\documentclass[a4paper,11pt]{article}
\pdfoutput=1 
             
\usepackage{jheppub} 
                     
\usepackage[T1]{fontenc} 
\usepackage{xspace}
\usepackage{tabularx}

\newcommand{\Bi}[1]{\ensuremath{^{#1}\mathrm{Bi}}\xspace}

\newcommand{\Pb}[1]{\ensuremath{^{#1}\mathrm{Pb}}\xspace}
\newcommand{\Po}[1]{\ensuremath{^{#1}\mathrm{Po}}\xspace}

\newcommand{\Rn}[1]{\ensuremath{^{#1}\mathrm{Rn}}\xspace}
\newcommand{\Th}[1]{\ensuremath{^{#1}\mathrm{Th}}\xspace}
\newcommand{\Tl}[1]{\ensuremath{^{#1}\mathrm{Tl}}\xspace}
\newcommand{\U}[1]{\ensuremath{^{#1}\mathrm{U}}\xspace}
\newcommand{\Xe}[1]{\ensuremath{^{#1}\mathrm{Xe}}\xspace}

\newcommand{\Kr}{$^{83\textrm{m}}$Kr\xspace}

\newcommand{\bbnonu}{\ensuremath{0\nu\beta\beta}\xspace}
\newcommand{\bbtwonu}{\ensuremath{2\nu\beta\beta}\xspace}
\newcommand{\bb}{\ensuremath{\beta\beta}\xspace}
\newcommand{\Qbb}{\ensuremath{Q_{\beta\beta}}\xspace}
\newcommand{\Qb}{\ensuremath{Q_{\beta}}\xspace}

\title{Radon-induced backgrounds in the NEXT-100 experiment}

\collaboration{The NEXT Collaboration}
\author[16]{C.~Cortes-Parra,}
\author[16]{P.~Novella,}
\author[16]{G.~Mart\'inez-Lema,}

\author[21]{H.~Almaz\'an,}
\author[23]{V.~\'Alvarez,}
\author[7]{L.~Arazi,}
\author[17]{I.J.~Arnquist,}
\author[20]{F.~Auria-Luna,}
\author[16]{S.~Ayet,}
\author[22]{Y.~Ayyad,}
% ...
\affiliation[7]{
Unit of Nuclear Engineering, Faculty of Engineering Sciences, Ben-Gurion University of the Negev, P.O.B. 653, Beer-Sheva, 8410501, Israel}
% ...
\affiliation[16]{
Instituto de F\'isica Corpuscular (IFIC), CSIC \& Universitat de Val\`encia, Calle Catedr\'atico Jos\'e Beltr\'an, 2, Paterna, E-46980, Spain}
\affiliation[17]{
Pacific Northwest National Laboratory (PNNL), Richland, WA 99352, USA}
% ...
\affiliation[20]{
Department of Organic Chemistry I, Universidad del Pais Vasco (UPV/EHU), Centro de Innovaci\'on en Qu\'imica Avanzada (ORFEO-CINQA), San Sebasti\'an / Donostia, E-20018, Spain}
\affiliation[21]{
Donostia International Physics Center, BERC Basque Excellence Research Centre, Manuel de Lardizabal 4, San Sebasti\'an / Donostia, E-20018, Spain}
\affiliation[22]{
Instituto Gallego de F\'isica de Altas Energ\'ias, Univ.\ de Santiago de Compostela, Campus sur, R\'ua Xos\'e Mar\'ia Su\'arez N\'u\~nez, s/n, Santiago de Compostela, E-15782, Spain}
\affiliation[23]{
Instituto de Instrumentaci\'on para Imagen Molecular (I3M), Centro Mixto CSIC - Universitat Polit\`ecnica de Val\`encia, Camino de Vera s/n, Valencia, E-46022, Spain}

\emailAdd{camilo.cortes@ific.uv.es}
\emailAdd{pau.novella@ific.uv.es}

\abstract{The NEXT-100 detector at the LSC aims at the first competitive search for the \bbnonu decay using a high-pressure \Xe{136} electroluminescent time projection chamber. The first low-background run of NEXT-100 at 3.95 bar has been devoted to the measurement of the radon-induced backgrounds impacting this search. The contributions from both the internal and external airborne radon have been evaluated. The internal \Rn{222} activity is found to be (0.95$\pm$0.04(stat)$\pm$0.09(sys)) Bq/m$^3$, while no traces of \Rn{220} have been observed. Most of the \Rn{222} progeny plate-out on the surface of the cathode of the detector, leading to a rate of Rn-induced \Bi{214} of (0.97$\pm$0.05(stat)$\pm$0.10(sys)) Hz for visible energies above 400 keV. The corresponding background index in the \bbnonu region of interest is evaluated as (7.3$\pm$1.5(stat)$\pm$0.8(sys))$\times10^{-4}$ counts/(keV$\cdot$kg$\cdot$yr) after selection of the fully contained events. This background index is reduced to $\sim$4$\times10^{-5}$ counts/(keV$\cdot$kg$\cdot$yr) by applying a topological selection requiring only one double-electron-like track in the events. This value is one order of magnitude below the total radiogenic background expectation in NEXT-100. By analyzing the correlation of the airborne radon activity and the measured rate of events in NEXT-100, it is concluded that the detector operates in a virtually radon-free environment thanks to the radon abatement system of the LSC.}

\begin{document} 
\maketitle
\flushbottom

\section{Introduction}
\label{sec:intro}

Revealing the nature of neutrino masses is one of the major goals in particle physics since oscillation experiments demonstrated that neutrinos are massive and that flavour number is not conserved. The observation of the neutrinoless double beta decay stands as the main probe to establish whether neutrinos are Majorana particles (fermions equivalent to their antiparticles), regardless of the underlying decay mechanism \cite{Gomez-Cadenas:2023vca}. Furthermore, this observation would provide insights into the absolute mass scale of these particles. Even-even nuclei can undergo double beta decay (\bb), a second order transition in which two bound neutrons transform simultaneously into two protons plus two electrons, when the beta decay is highly suppressed or energetically forbidden. The \bb mode in which two antineutrinos are emitted (\bbtwonu) has been observed in ten nuclides with half-lifes in the range of $\sim$$10^{19}$–$10^{21}$ yr \cite{Barabash:2020nck}. The mode with no neutrinos emitted (\bbnonu), which violates the lepton number by two units, has not been observed, and the best current limits to the half-life of this process have been reported to exceed 10$^{26}$ yr \cite{KamLAND-Zen:2024eml}.


The NEXT experiment is searching for the \bbnonu process using high-pressure \Xe{136}-gas electroluminescent time projection chambers (TPCs). The NEXT-White detector \cite{NEXT:2018rgj} was the first radiopure implementation of this technology operating underground in the Laboratorio Subterr\'aneo de Canfranc (LSC). NEXT-White achieved an excellent energy resolution of $\sim$1\% FWHM at 2.6 MeV \cite{Renner:2019pfe} (above the \Qbb of \Xe{136}, (2.4578$\pm$0.0004) MeV \cite{Redshaw:2007un}), and demonstrated efficient background rejection factors by means of the topological information of the TPC \cite{NEXT:2019gtz,NEXT:2020jmz,NEXT:2020try}. In addition to the characterization of the \bb backgrounds associated to the NEXT technology \cite{Novella:2018ewv,Novella:2019cne}, this demonstrator provided also the first measurement of the half-life of the \bbtwonu decay in gaseous xenon \cite{NEXT:2021dqj}, as well as a first search for the \bbnonu process yielding a limit in the order of $10^{24}$ yr at 90\% C.L. \cite{NEXT:2023daz}. After the successful exploitation of NEXT-White, the NEXT-100 detector \cite{NEXT:2025yqw} has been installed in the HALL A of the LSC and is in operation since 2024. Holding up to 100 kg of xenon at 15 bar, the main goals of NEXT-100 are to perform a first competitive search for the \bbnonu in gaseous xenon, reaching sensitivities in the order of $10^{25}$ yr at 90\% C.L. after 3 years of operation \cite{NEXT:2015wlq}, and to serve as a test-bench for the technological solutions to be implemented in future ton-scale detectors \cite{NEXT:2020amj}. The first physics run of NEXT-100, operating at $\sim$4 bar, has been devoted to the characterization of the detector performance in terms of energy resolution \cite{NEXT:2025fpq,NEXT:2025ozn}, as well as to the measurement of the radiogenic backgrounds in order to ensure the radiopurity of the device. Among the different background contributions, the present work focuses on the Rn-induced ones.     


\begin{figure}
  \begin{center}
    \includegraphics[scale=0.23]{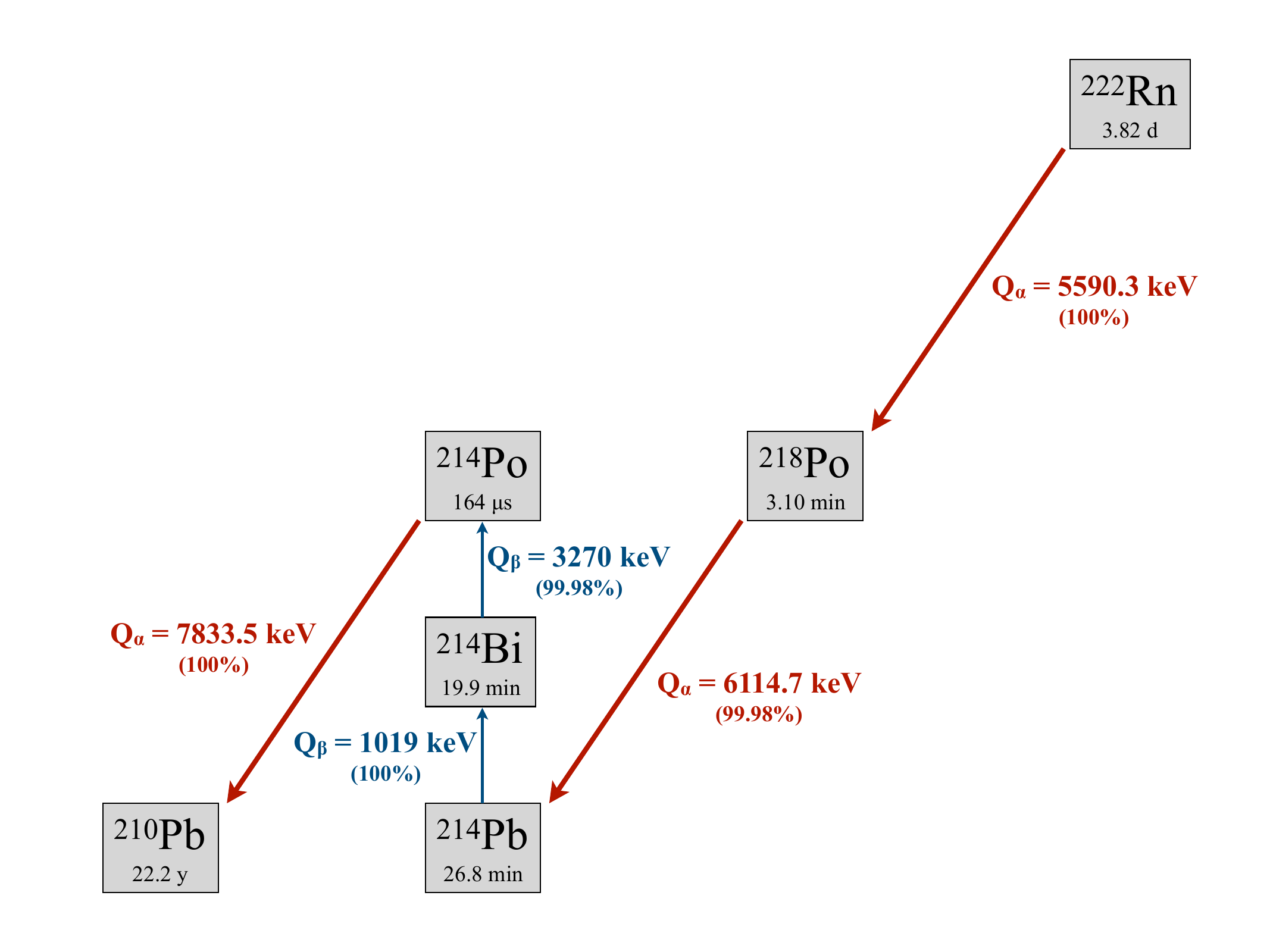}
    \includegraphics[scale=0.23]{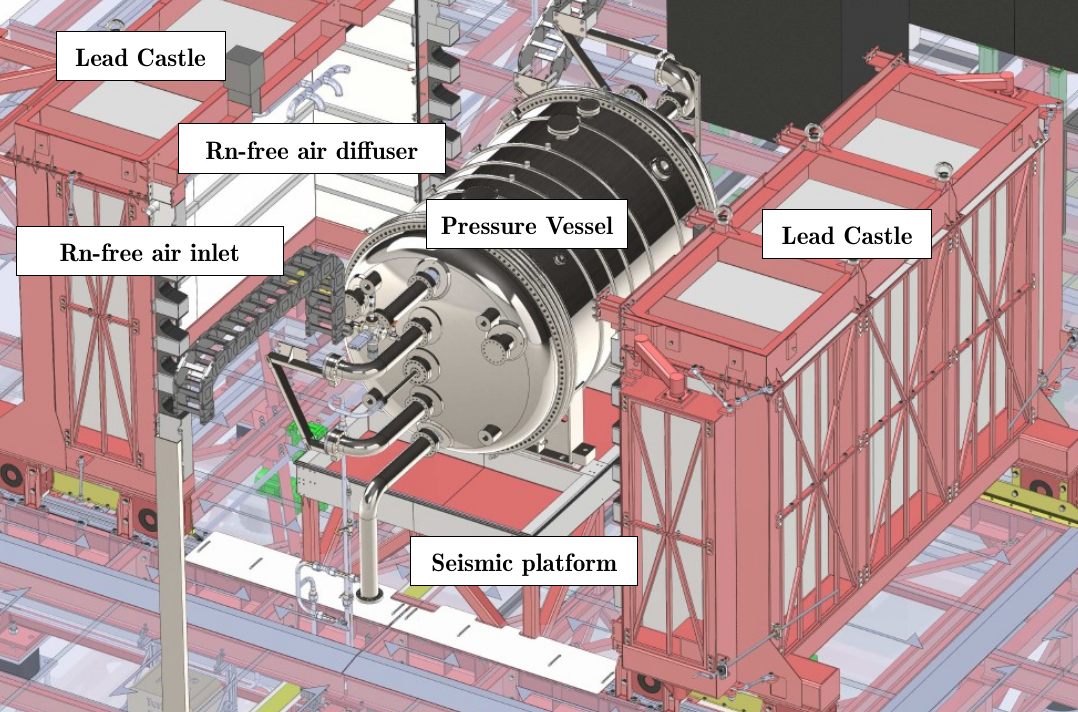}
    \caption{Left: \Rn{222} decay chain ending in the long-lived \Pb{210} isotope. Quoted half-lives,\linebreak Q-values and branching ratios are extracted from the ENSDF database \cite{Wu:2009lpp,Jain:2006yav,ShamsuzzohaBasunia:2014yyr}. Right: render of the NEXT-100 detector with the lead castle in the open position and the related LSC facilities.}
    \label{fig:rndecay}
  \end{center}
\end{figure}

\Rn{220} and \Rn{222} from the naturally-occurring \Th{232} and \U{238} decay chains, respectively, can become one of the most relevant backgrounds in the search for rare events, such as the neutrinoless double beta decay or dark matter interactions. The decays of \Rn{222} and its progeny, shown in left panel of Fig.~\ref{fig:rndecay}, are of particular interest for \bbnonu experiments due to the high-energy gammas (up to 3.18 MeV \cite{Wu:2009lpp}) involved in the \Bi{214} $\beta$ decay. Radon is a highly diﬀusive and soluble noble gas. \Rn{222}, with a relatively long half-life of 3.82 day \cite{Jain:2006yav}, can emanate from the surface of the materials where it was originally trapped. This applies to both detector materials and the rock of the walls of the underground laboratories, becoming a background with a variable spatial distribution and concentration, depending on the detector and laboratory conditions (like temperature or pressure). The daughter isotopes produced in the radon decay (\Po{218} and \Pb{214}) are often positively-charged ions, and thereby tend to plate-out on surfaces where an electrostatic field is present. This leads to eventual concentrations of \Bi{214} on those charged surfaces (see for instance \cite{Novella:2018ewv,NEMO:2009ewu}). Overall, Rn-induced backgrounds arising from the outgassing of \Rn{222} require to be treated in dedicated ways, differing from those addressing radio-impurities trapped in detector or laboratory materials. In particular, \bbnonu experiments need to monitor and mitigate both internal (from inner detector components) and airborne (from the atmosphere surrounding the detector) Rn-induced backgrounds. In this work, these internal and external background sources in the NEXT-100 experiment are analyzed.       

This paper is organized as follows: Sec.~\ref{sec:next100} describes the NEXT-100 detector, as well as the related infrastructures in the LSC. Sec.~\ref{sec:lpr} presents the operation conditions of the detector, and the dedicated data-taking campaigns to measure and characterize the \linebreak Rn-induced backgrounds. Finally, Sec.~\ref{sec:internal} and Sec.~\ref{sec:airborne} study the internal and airborne contributions to the background budget of NEXT-100, respectively, evaluating their impact on the \bbnonu search.

\section{The NEXT-100 detector}
\label{sec:next100}

A general description of NEXT-100 and its auxiliary subsystems has been presented in \cite{NEXT:2025yqw}.\linebreak The detector exploits the primary scintillation light (S1) and the ionization electrons generated by charged particles in xenon. The core of NEXT-100 is the TPC, which is designed to operate with xenon gas at a maximum pressure of 15 bar ($\sim$100 kg) inside a large pressure vessel made of radiopure stainless steel (316Ti). The TPC consists of three main axial sections: 1) drift region ($\sim$0.901 m$^3$): extends between the cathode and a gate meshes, maintaining a uniform electric field where ionization electrons drift toward the anode, 2) electroluminescence (EL) region ($\sim$0.009 m$^3$): a small gap of (9.70$\pm$0.15) mm between the gate and anode where a strong electric field accelerates the drifting electrons to produce a secondary scintillation light signal (S2) \cite{NEXT:2023blw}, which is used for energy measurement and tracking, and 3) buffer region ($\sim$0.183 m$^3$): located near the energy plane, between the cathode and the inner copper shielding, to prevent sparking and unwanted electroluminescence. The TPC is contained within a Field Cage (FC), made of 18 HDPE struts and 52 copper rings to shape the electric field along the drift region. The active volume is further defined by PTFE reflector panels coated with tetraphenyl butadiene (TPB), a wavelength shifter, to create a light-tube that enhances photon collection efficiency.

The NEXT-100 detector uses an asymmetric design with two independent sensor planes on opposite ends of the TPC to collect the S1 and S2 light signals. The energy plane (EP), located behind the cathode, measures the initial time of the events (S1) and their energy~(S2). It is instrumented with up to 60 Photomultiplier Tubes (PMTs) (R11410-10 from Hamamatsu), although during the first run of the detector only 53 have been installed. The PMTs are coupled to the xenon gas volume through sapphire windows coated with PEDOT and a thin layer of TPB. The tracking plane (TP), placed in front of the EL region, provides spatial and topological information of the event for 3D reconstruction. It consists of a matrix of 3,584 Silicon Photomultipliers (SiPMs) (S13372-1350TE from Hamamatsu) placed in contact with the xenon gas. The SiPMs are distributed on 56 Kapton PCBs of 8$\times$8 SiPMs, which are covered by PTFE masks coated with TPB.

The detector is located in the HALL A of the LSC (overburden of 2450 m of water equivalent), providing significant shielding from cosmic radiation. In addition, NEXT-100 is protected from the environmental backgrounds by a two-layer shielding system: 1) an outer lead castle (LC) made of bricks with a total thickness of 200 mm to reduce the impact of the external radioactivity, and 2) an inner copper shielding (ICS) made of ultra-pure copper bars with a thickness of 120 mm and installed inside the pressure vessel, meant to shield the radioactivity from both the pressure vessel and the LC itself. Outside the LC, a gas system is continuously delivering purified xenon to the detector, via recirculation through chemical purifiers: two ambient temperature getters (SAES MicroTorr MC4500-902FV) and one heated getter (SAES MonoTorr PS4-MT50-R-535). As the radon emanation from the cold getters is too high for the \bbnonu searches (see Ref.~\cite{Novella:2018ewv}), only the hot getter is used when the detector is operating in low-background mode. The cold getters are used instead for commissioning or calibration purposes (see Ref.~\cite{NEXT:2025yqw}), as well as to characterize the Rn-induced background as shown in this work. In order to reduce the airborne radon concentration in the air volume enclosed by the LC, a radon abatement system (RAS) by ATEKO A.S. delivers virtually radon-free air into this volume \cite{Perez-Perez:2021bzu}. With a ﬂow from 180 m$^3$/h to 220 m$^3$/h, the air delivered by the RAS pipes reaching the lead castle has a radon concentration of about 1 mBq/m$^3$, which implies a reduction of 4-5 orders of magnitude with respect to HALL A air. The right panel of Fig.~\ref{fig:rndecay} shows a render of the NEXT-100 detector and the associated LSC facilities, including the inlet and diffuser for the RAS air delivery.    


\section{Low-pressure run operation conditions}
\label{sec:lpr}

In order to assess the Rn-induced backgrounds in NEXT-100, two dedicated data-taking campaigns have been conducted, operating the detector with xenon depleted in the \Xe{136} isotope: a radon-background run and a low-background run. The first one is devoted to the measurement of the internal radon backgrounds, collecting alpha decays within the active volume. The second one is meant to evaluate the impact of the airborne radon in the air enclosed by the lead castle, measuring the electron-like backgrounds. In both runs, the operation pressure has been kept constant at $\sim$3.95 bar. While some operation conditions and the trigger definition are different in both periods, the data acquisition system (DAQ) is common and delivers for each registered event the 53 PMT waveforms sampled at 25 ns, and the 3,584 SiPMs waveforms sampled at 1 $\mu$s. Tab. \ref{tab:data} summarizes the different data samples described in this section.

\begin{table}[!htb]
\caption{\label{tab:data} Data-taking periods and corresponding detector operation conditions.}
\begin{center}
\resizebox{\textwidth}{!}{
\begin{tabular}{c|cccccc}
\hline
Period & Trigger type &  Drift field (V/cm) & EL field (kV/cm) & Getter & Lead castle & RAS  \\
\hline
A1 & $\alpha$ & 130 & 7.7 & Cold & Open & Off \\
A2 & $\alpha$ & 130 & 7.7 & Hot  & Open & Off \\
\hline
E1 & e$^-$ & 120 & 9.1 & Hot & Closed & Off \\
E2 & e$^-$ & 120 & 9.1 & Hot & Closed & On  \\
\hline
\end{tabular}
}
\end{center}
\end{table}

\subsection{Radon-background run}
\label{sec:rnrun}

\begin{figure}
  \begin{center}
    \includegraphics[scale=0.35]{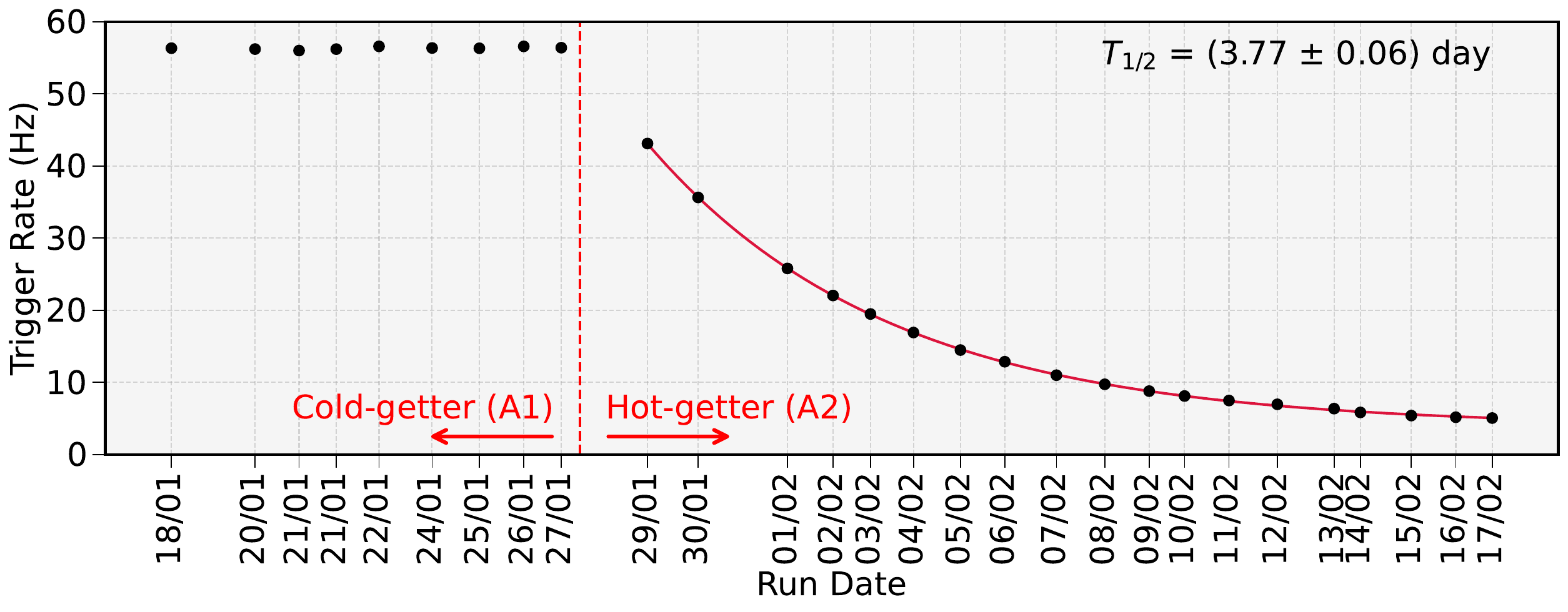}
    \caption{ Dead-time-corrected trigger rate as function of the date of data-taking, during the A1 (high radon activity) and A2 (low radon activity) periods. The beginning of A2, corresponding to the start of the xenon recirculation through the hot getter, is marked with a dashed red line. The solid red line in A2 corresponds to a fit yielding a \Rn{222} half-life of 3.77$\pm$0.06 day. Error bars include both statistical and uncorrelated systematic uncertainties.}
    \label{fig:alphatrigger}
  \end{center}
\end{figure}

To determine the contribution of internal radon, dedicated runs triggering on alpha decays have been taken without using any external calibration source. To avoid sensor saturation, the EL field is limited to $\sim$7.7 kV/cm (7.5 kV at the gate), while the drift field is set to $\sim$130 V/cm (23 kV at the cathode). The corresponding electron drift velocity is around 0.88mm/$\mu$s. The S2 signals recorded in one of the central PMTs are used to trigger the DAQ system, by imposing a minimum value on the integrated charge of 10,000 ADC and a time width between 2 $\mu$s and 400 $\mu$s. For each trigger, a readout buffer of 2.6 ms is considered, with a pre-trigger window of 1.6 ms allowing to register the S1 signal associated to the event. The alpha data-taking has consisted of two different periods: a high radon activity run (A1) recirculating the xenon gas through the cold getters in the gas system (from which large amounts of $^{222}$Rn emanate), and a low radon activity run (A2) purifying the gas through the hot getter. While the A1 data are meant to characterize with large statistical samples the distribution and energy spectrum of the alpha decays, the A2 sample is used to measure the intrinsic (irreducible) level of internal Rn-induced background in NEXT-100, whose origin is the radon emanation from the detector and gas system surfaces. According to the different total trigger rates, the DAQ dead-time has ranged from $\sim$48\% at the beginning of A1, to $\sim$8\% at the end of A2. The DAQ dead-time-corrected alpha trigger rate as a function of the date of data-taking is shown in Fig.~\ref{fig:alphatrigger}. The decrease in rate over time during A2 is a reflection of the half-life of $^{222}$Rn, while the trigger rate at the end of this period can be associated with the irreducible level of $^{222}$Rn within the NEXT-100 detector (see Sec.~\ref{sec:internal}). The trigger rate has been fitted to an exponential function plus a constant term. While the former accounts for the decay of the \Rn{222} released by the cold getter, the latter reflects the intrinsic level of \Rn{222} during the operation with the hot getter, which is assumed to be constant in time. The fit considers an uncorrelated systematic uncertainty accounting for variations in the alpha rate during A1 (0.5\%). The fit results in a half-life of (3.77$\pm$0.06) day, fully consistent with the accepted value for $^{222}$Rn of (3.8235$\pm$0.0003)~day~\cite{Jain:2006yav}, and an intrinsic constant rate of (3.89$\pm$0.17)~Hz. This latter value can be compared with the trigger rate during the A1 period, which was measured to be (56.36$\pm$0.28) Hz.

\subsection{Low-background run}
\label{sec:lowbgrun}

\begin{figure}
  \begin{center}
    \includegraphics[width=\columnwidth]{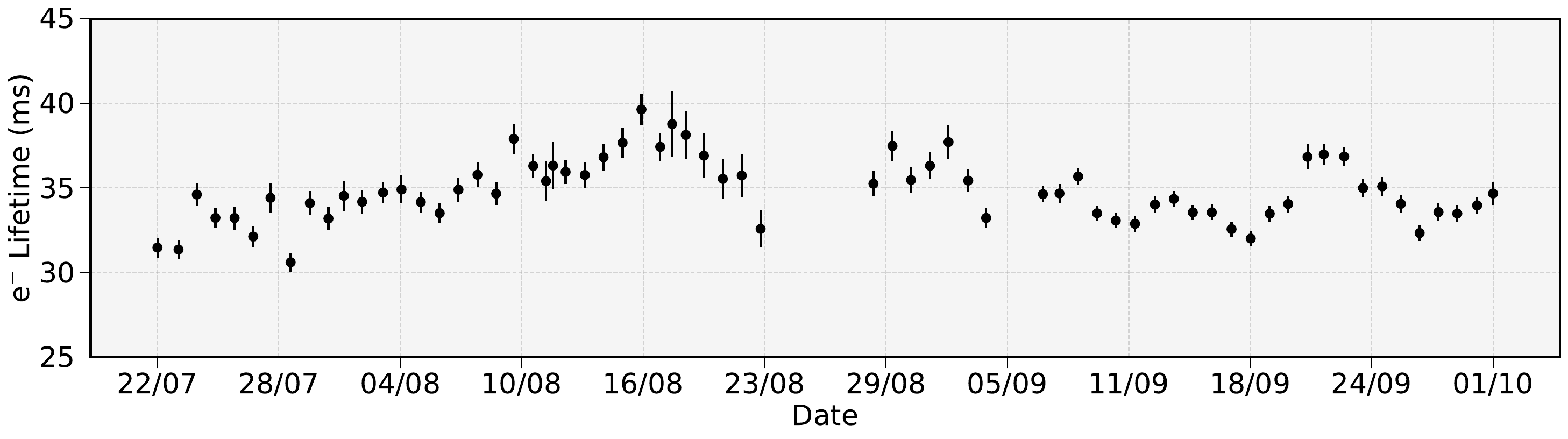}
    \caption{ Electron lifetime evolution during the low-background data-taking period during 2025, as derived from the \Kr calibration. An average value for each DAQ run ($\sim$24 hours) is shown.}
    \label{fig:lifetime}
  \end{center}
\end{figure}

The impact of the airborne radon in the air inside the lead castle, in its closed position, has been evaluated with a first low-background run triggering on electron events, adopting the typical operation and trigger conditions used in the data-taking for the \bbnonu search. The drift and the EL fields have been set to $\sim$120 V/cm (23 kV at cathode) and $\sim$9.1~kV/cm (8.8 kV at gate), respectively. These values correspond to the maximum fields at which the detector has been operated in a stable manner, in particular avoiding sparks. Under these conditions, the drift velocity has been kept at $\sim$0.87 mm/$\mu$s. The dual-trigger capabilities of the DAQ system \cite{NEXT:2025yqw} have been exploited in order to collect both low-energy (below $\sim$100 keV) and high-energy (above $\sim$400 keV) triggers within the same run. While the low-energy trigger (T1) is devoted to \Kr events used to continuously calibrate the detector \linebreak(see Ref.~\cite{NEXT:2025fpq}), the high-energy trigger (T2) collects the backgrounds impacting the \bb searches. S2-based trigger conditions have been defined for each one considering one of the central PMTs. For T1, an integrated charge between 1600 ADC and 7000 ADC and a time width between 0.5 $\mu$s and 20 $\mu$s are required. For T2, signals must have a minimum integrated charge of 30,000 ADC and a minimum time width of 2 $\mu$s. The corresponding T1 and T2 DAQ dead-times are around 20\% and 25\%, respectively. These large DAQ inefficiencies are  mostly due to the high Kr rate ($\sim$30 Hz) and  will be significantly reduced when collecting data for the \bbnonu search by limiting the throughput of the T1 data stream. The xenon gas has been recirculated continuously through the hot getter, ensuring a minimal level of internal \Rn{222} while keeping a high electron lifetime. According to the analysis of the \Kr data, the lifetime has ranged between $\sim$30 ms and $\sim$40 ms as shown in Fig.~\ref{fig:lifetime}, which imply corrections in the reconstructed charge below 5\% for the maximum drift distance. Although the decrease with respect to the electron lifetime in the A1 and A2 periods is not fully understood, it might be explained by the introduction of impurities from the \Kr bypass in the gas system. The energy scale has been stable, with variations below 2\%. The low-background run has in turn been divided into two periods: while the radon abatement system was not in operation during the first one (E1), it was in continuous operation during the second (E2). As shown in Sec.~\ref{sec:airborne} the analysis and comparison of both data samples has allowed for an assessment of the impact of the airborne radon in NEXT-100.

\section{Internal Rn-induced background}
\label{sec:internal}
The activity of the internal radon, from outgassing of detector materials, as well as the impact of the Rn-induced backgrounds in the \bbnonu search, are derived from the data samples collected in the A1 and A2 periods. First, the internal radon activity and the induced contamination of \Bi{214} on the cathode surface are estimated by analyzing the energy spectrum of the alpha decays within the \Rn{222} decay chain (Secs.~\ref{sec:alphareco}-\ref{sec:bicath}). Then, the remaining Rn-induced \Bi{214} background is computed after the typical selection cuts applied to select \bb candidate events (Sec.~\ref{sec:bb0nubg}).      

\subsection{Reconstruction and selection of $\alpha$ particles}
\label{sec:alphareco}

\begin{figure}
  \begin{center}
    \includegraphics[scale=0.30]{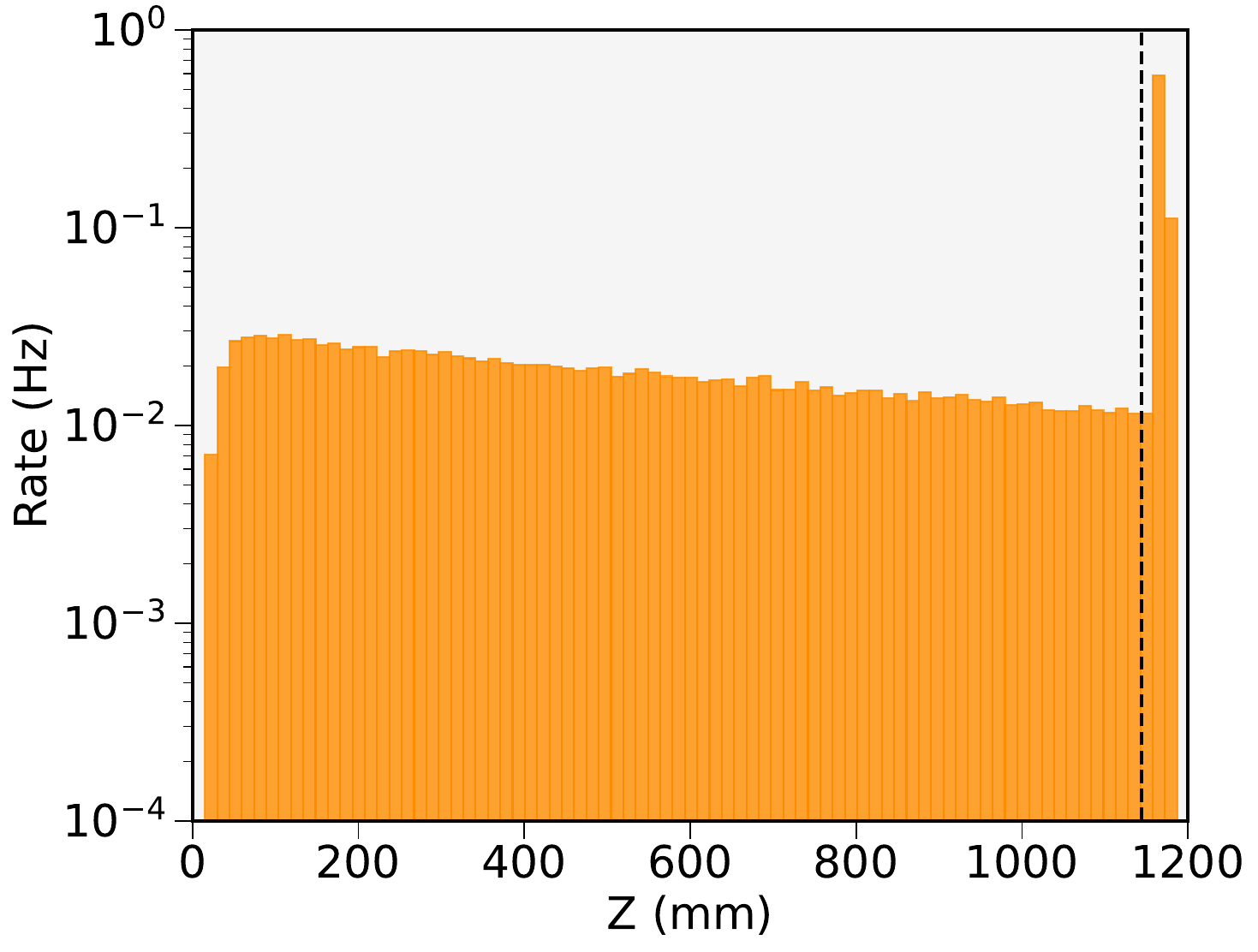}
    \includegraphics[scale=0.325]{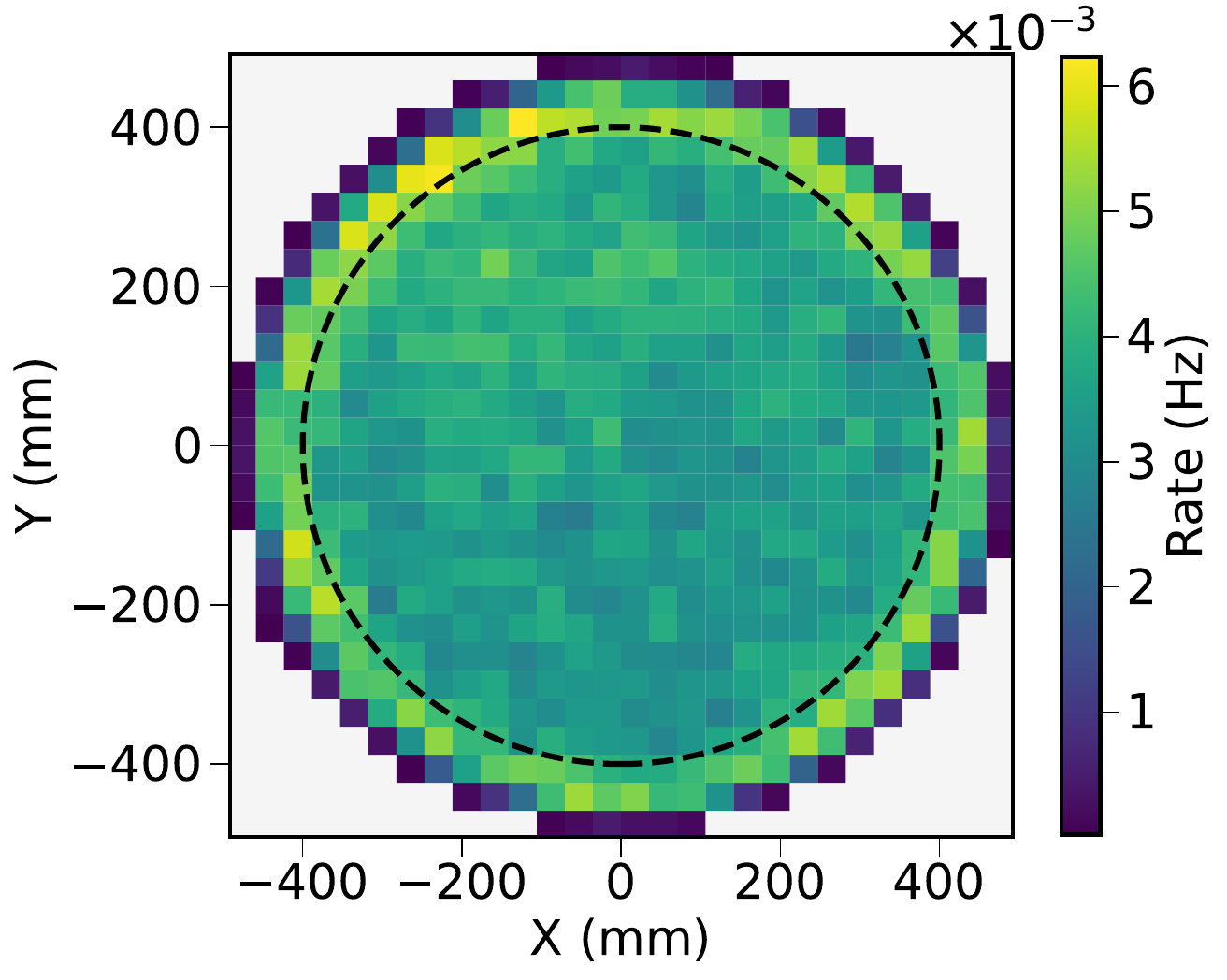}
    \caption{Spatial distribution of alpha candidate events. Left panel: Z distribution. Right panel: distribution in the (X,Y) plane. The dashed black lines indicate the fiducial region.}
    \label{fig:aphadist}
  \end{center}
\end{figure}

The reconstruction of the alpha events is performed in two steps. First, S1 (time width below 4 $\mu$s) and S2 (time width above 10 $\mu$s) signals are reconstructed within a global waveform obtained by summing up the individual PMT waveforms. Then, a point-like position is reconstructed by considering the SiPM hits in the time window of the identified S2 signal. The point-like reconstruction is a reasonable approximation provided that the Continuous Slowing Down Approximation (CSDA) range of an alpha particle of 5.5 MeV is 45 mm at 4 bar. The XY coordinates in the tracking plane is determined from the barycenter of the charge registered by the individual SiPMs, while the Z position along the drift axis is obtained from the time difference between the S1 and S2 signals and the drift velocity. From the reconstructed events, alpha candidates are selected by requiring only one S1 and only one S2 signals in the PMT waveforms. The spatial distribution of inclusive alpha candidates during the A1 period is shown in Fig.~\ref{fig:aphadist}. The distribution along the Z axis shows a clear peak at the cathode position ($\sim$1187 mm). As reported in Ref.~\cite{Novella:2018ewv}, the  radon progeny, mostly generated in ionized form, plate-out on the cathode surface. The distribution in the XY plane is quite uniform up to a radial distance of $\sim$400 mm, showing an increase of alpha events in the active volume boundaries. In order to select alpha decays in the xenon gas, a fiducial volume of Z$<$1143 mm and R=$\sqrt{\textrm{X}^2+\textrm{Y}^2}$<400~mm is considered, as shown with dashed black lines in Fig.~\ref{fig:aphadist}. Although with a negligible impact in the current analysis, events not properly reconstructed are filtered out according to a shape analysis of the signals.

The energy estimator for the alpha decays is built along the lines of Ref.~\cite{Novella:2018ewv,NEXT:2014ays}. First, the raw energy determined from the integral of the S2 signal is corrected by the XY inhomogeneities by means of an energy response map generated with an independent statistical sample of alpha events. In addition, a Z-dependent S2 correction accounting for the ionization electron attachment is applied. The electron lifetime has been around 50~ms during both periods, with $\sim$20\% variations that do not imply a significant change in the S2 light yield. The integral of the S1 signal is also corrected by the variation of the light collection along the Z axis. Then, in order to reduce the impact of the recombination fluctuations in the energy resolution, the energy estimator is defined as a linear combination of the primary and secondary scintillation light:

\begin{equation}
E\equiv \lambda (E_{S1}+E_{S2}/\eta)
\label{eq:es1s2}
\end{equation}

\noindent where $E_{S1}$ and $E_{S2}$ are the corrected yields for the S1 and S2 signals, respectively, $\eta$ is a weight factor to rescale the ionization yield component, and $\lambda$ is an overall energy scale factor. The value of $\eta$ is chosen so it provides the best energy resolution in the \Rn{222} peak. The energy scale factor, providing the conversion between the yield in p.e and in keV, is determined by aligning the fitted \Rn{222} peak position with 5590.3 keV.

\subsection{Measurement of the internal radon activity}
\label{sec:rnactivity}

The last five days of the A2 period are considered in the determination of the intrinsic radon contamination, as they reflect to a good approximation the data-taking conditions when purifying the xenon gas through the hot getter. The quoted trigger rates in Fig.~\ref{fig:alphatrigger} account for some fraction of events not induced by alpha decays. This fraction has been evaluated by analyzing a sample of PMT waveforms and counting the number of events not containing an alpha-like S1 signal. The fraction of triggered events corresponding to alpha decays was measured to be (75.4$\pm$7.5)\% for the A1 period and (30.6$\pm$3.1)\% for the A2 period. A conservative systematic uncertainty of 10\% has been assigned to cover the variations in the light collection efficiency across the active volume. Accordingly, the alpha trigger decay rates in the high and low radon activity data samples are R$^{\text{trg}}_{\textrm{A1}}$($\alpha$)=(42.5$\pm$0.2(stat)$\pm$4.2(sys))~Hz and R$^{\text{trg}}_{\textrm{A2}}$($\alpha$)=(1.19$\pm$0.05(stat)$\pm$0.12(sys))~Hz, respectively.

\begin{figure}
  \begin{center}
    \includegraphics[scale=0.30]{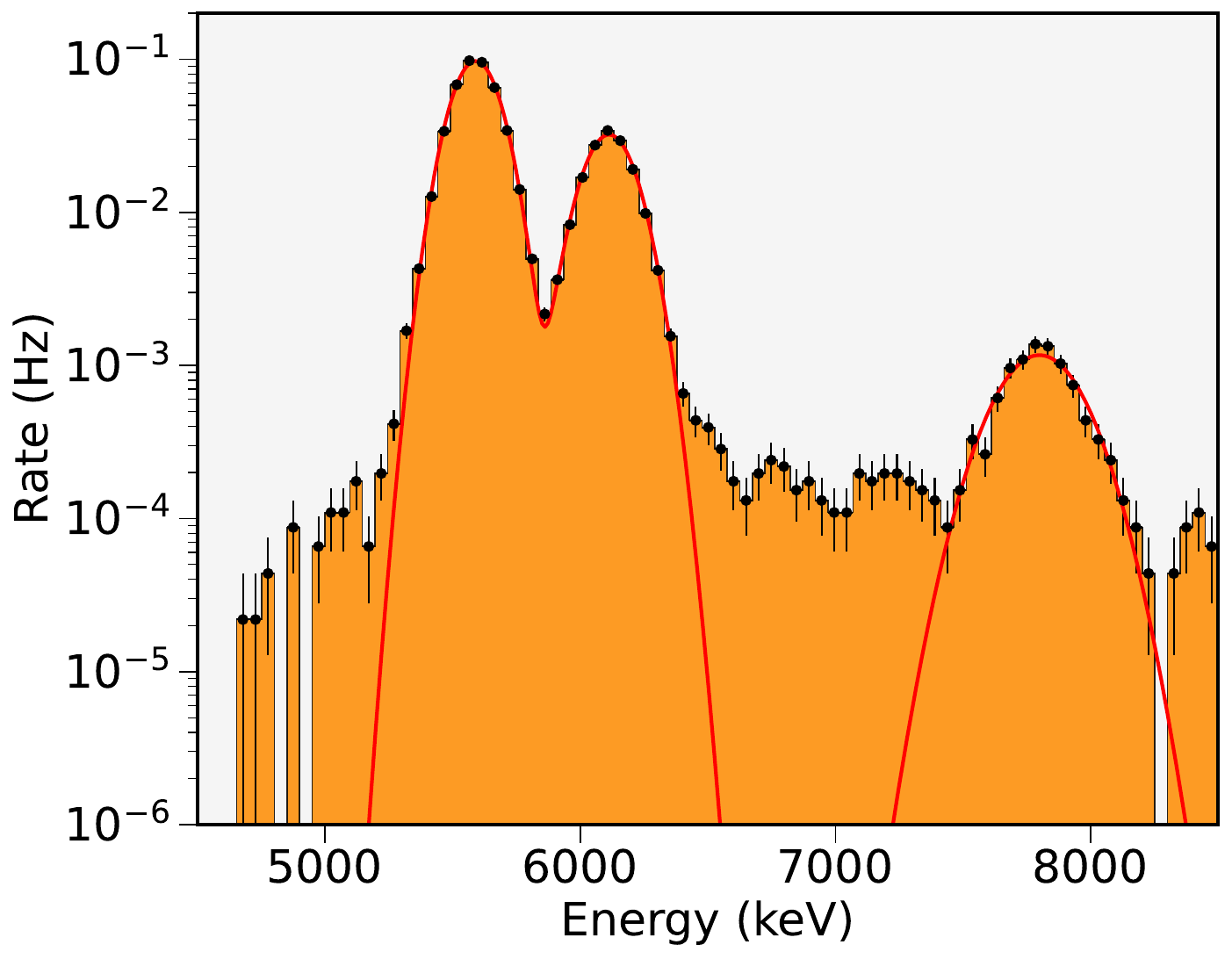}
    \includegraphics[scale=0.30]{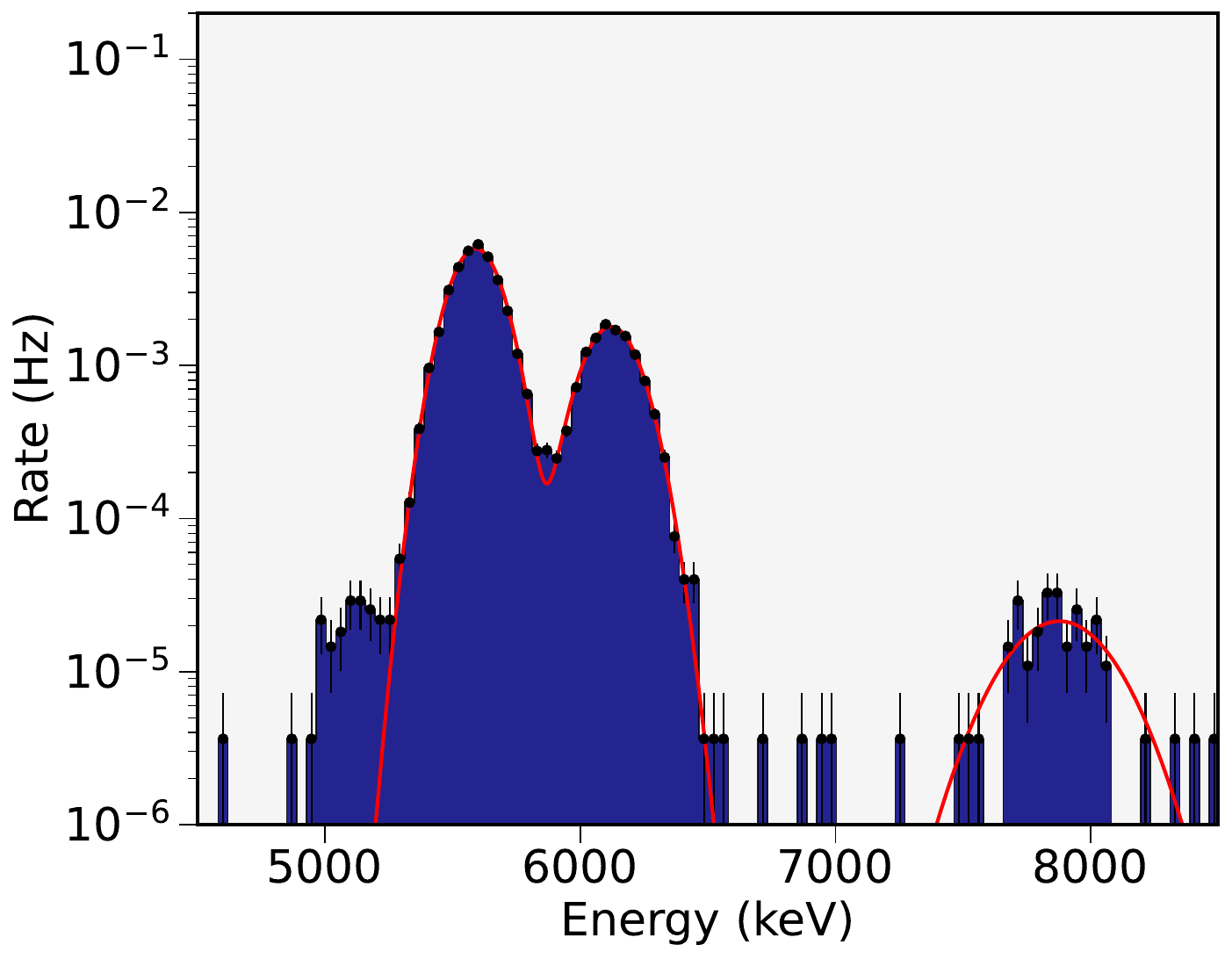}
    \caption{Energy distribution for fiducial alpha candidate events during A1 (left) and A2 (right) periods. A triple Gaussian fit is superimposed to describe the \Rn{222} (5590~keV), \Po{218} (6115~keV) and \Po{214} (7834~keV) populations.}
    \label{fig:alphaspec}
  \end{center}
\end{figure}

Although the time evolution of the trigger rate and the derived half-life value confirms that the observed alpha decays are due to radon and its progeny, the determination of the relative contributions of the \Rn{222} (5590~keV), \Po{218} (6115~keV) and \Po{214} (7834~keV) isotopes (see Fig.~\ref{fig:rndecay}) is performed by means of a spectroscopic analysis of the fiducial events. After applying the reconstruction and selection process previously described, the resulting rates in the high and low radon periods are R$^{\text{sel}}_{\textrm{A1}}$($\alpha$)=(610.9$\pm$3.7) mHz and R$^{\text{sel}}_{\textrm{A2}}$($\alpha$)=(49.3$\pm$0.4) mHz, respectively. The energy spectra of the events collected in the A1 and A2 periods have been fitted to the sum of three Gaussian functions, as shown in Fig.~\ref{fig:alphaspec}. Despite some non-Gaussian tails due to the inaccuracy in the energy reconstruction of some events, the populations of the \Rn{222}, \Po{218} and \Po{214} are clearly visible. No traces of other decays (like \Rn{220} and its progeny) are observed. From the fit results, the corresponding decay rates of the three isotopes are computed and listed in Tab.~\ref{tab:rates}. The total rate extracted from the fit is also in reasonable agreement (within 3\%) with the total rate of selected events, thus demonstrating that the events in the data samples are well described by the \Rn{222} decay chain. The \Rn{222} decays represent a (70.8$\pm$0.7)\% and (72.1$\pm$0.9)\% fraction of the alpha decays in the A1 and A2 periods, respectively. Taking into account the total alpha trigger rates, R$^{\text{trg}}$($\alpha$), and the active volume, the derived radon activities in both periods are A$_{\textrm{A1}}$(\Rn{222})=(33.4$\pm$0.4(stat)$\pm$3.3(sys))~Bq/m$^3$ and A$_{\textrm{A2}}$(\Rn{222})=(0.95$\pm$0.04(stat)$\pm$0.09(sys))~Bq/m$^3$.

\begin{table}[!htb]
\caption{\label{tab:rates} Best-fit alpha decay rates in the \Rn{222} decay chain.}
\begin{center}
\resizebox{\textwidth}{!}{ 
\begin{tabular}{c|ccc|c}
\hline
Period  & \Rn{222} Rate (mHz) & \Po{218} Rate (mHz)  & \Po{214} Rate (mHz) & Total Rate (mHz) \\ \hline
A1      & 432.5$\pm$3.1 & 156.2$\pm$1.8  & 9.1$\pm$0.4   & 597.8$\pm$3.6 \\
A2      & 35.6$\pm$0.4  & 12.2$\pm$0.2   & 0.27$\pm$0.04 & 48.1$\pm$0.4 \\
\hline
\end{tabular}
}
\end{center}
\end{table}

\subsection{Measurement of the Rn-induced \Bi{214} activity at the cathode}
\label{sec:bicath}


From the measurement of the internal radon activity, the Rn-induced \Bi{214} activity on the cathode surface when circulating the gas through the hot getter (standard operation when collecting data for the \bbnonu search) can be computed as described in Ref.~\cite{Novella:2018ewv}. According to Tab.~\ref{tab:rates}, in the low radon activity data the relative fractions of the \Po{218} and \Po{214} populations with respect to \Rn{222} are (34.2$\pm$0.7)\% and (0.76$\pm$0.11)\%, respectively. As the branching ratios of these decays are $\sim$100\%, the complementary fractions correspond to the ionized radon progeny which plates out on the cathode grid, thereby not decaying in the active volume. Although it is difficult to estimate the proportions of neutral and charged \Po{218} and \Po{214} atoms, it is expected that most of them are in ionized form in a gaseous detector \cite{NEMO:2009ewu}. On the contrary, the fraction of neutral atoms increases in a liquid device due to higher electron-ion recombination \cite{Albert:2015vma}. The fraction of the \Po{218} to \Po{214} alpha decay rates in the fiducial volume, F(\Po{214}/\Po{218}), is (2.2$\pm$0.3)\% in the A2 period (see Tab.~\ref{tab:rates}). The fraction of neutral \Pb{214} produced in the \Po{218} decay, F(\Pb{214}/\Po{218}), can be approximated to be the same as in the \Rn{222} decay to \Po{218}, F(\Po{218}/\Rn{222})=(34.2$\pm$0.7)\%, given the similarity of both alpha decays. Given the short half-life of \Po{214} (164 $\mu$s), these fractions are used to derive the fraction of neutral \Bi{214} isotopes as 

\begin{equation}
\textrm{F}(\Bi{214}/\Pb{214}) \simeq \frac{\textrm{F}(\Po{214}/\Po{218})}{\textrm{F}(\Pb{214}/\Po{218})}
\label{eq:bi214ratio}
\end{equation}

\noindent which yields (6.5$\pm$1.0)\%, where the quoted uncertainty is only statistical. Thus, it is concluded that $\sim$93\% of the Rn-induced \Bi{214} decays take place on the cathode surface. Given the measured activity of the intrinsic internal radon, A$_{\textrm{A2}}$, and the active volume, the rate of \Bi{214} events on the cathode is estimated to be (0.80$\pm$0.04(stat)$\pm$0.08(sys))~Hz. A minor contribution from the radon decays in the buffer region has to be added to this rate. According to the ratio of the active and buffer volumes, the expected \Bi{214} contamination due to buffer events is estimated as (0.17$\pm$0.06(stat)$\pm$0.02(sys)) Hz, for a total rate of R(\Bi{214})=(0.97$\pm$0.05(stat)$\pm$0.10(sys)) Hz.  

\subsection{Rn-induced background in $\beta\beta0\nu$ searches}
\label{sec:bb0nubg}

\begin{figure}
  \begin{center}
    \includegraphics[scale=0.40]{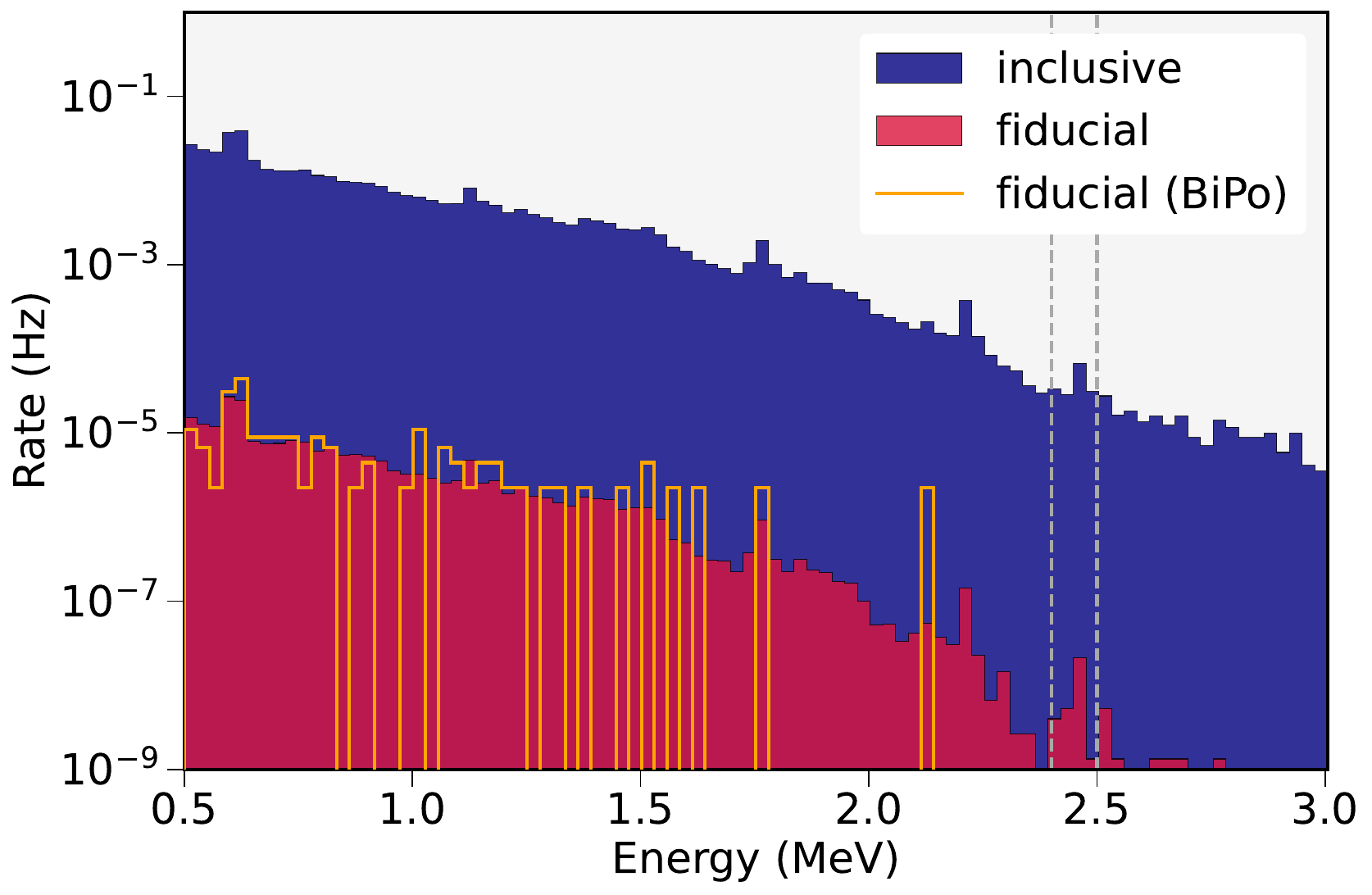}
    \caption{Rn-induced \Bi{214} background, before (solid blue histogram) and after (solid red histogram) applying the fiducial selection described in the text. The \Bi{214} PDF is normalized according to the corresponding rate of events in the statistical-limited MC sample simulating the \Bi{214}$\rightarrow$\Po{214} chain (solid yellow line). The dashed vertical lines mark the ROI around the \Qbb of \Xe{136} considered to compute the background rejection factors of the \bb-candidate selection.}
    \label{fig:bi214}
  \end{center}
\end{figure}

The decay of \Bi{214} (\Qb=3.27 MeV) can become a background in the \bbnonu search due to the high energy gammas involved, and in particular due to the 2.448 MeV gamma (with a branching ratio of 1.55\%), which is only 10 keV away from the \Qbb of \Xe{136}. However, NEXT detectors suppress this background very efficiently thanks to the associated $\beta$ particle, as well as to the $\alpha$ decay of \Po{214}.  The impact of the Rn-induced \Bi{214} contamination is evaluated by means of a MC simulation performed with \textsc{Nexus} \cite{NEXUS}, a GEANT4-based~\cite{Agostinelli:2002hh} software describing in detail the different components of the NEXT-100 detector and its operation conditions. The electron drift effects (including diffusion and electron attachment), the production of the scintillation light (S1 and S2) and its propagation, as well as the response of the photo-sensors in the energy and tracking planes are simulated. Two dedicated MC samples of \Bi{214} decays on the cathode have been used. The first one considers the full simulation of the \Bi{214}$\rightarrow$\Po{214} decay chain, with the proper time distribution of the $\beta/\gamma$ and $\alpha$ emissions. As this simulation is significantly CPU-consuming, a limited statistical sample ($\sim$275,000 reconstructed events) has been produced with the goal of assessing the background rejection factors of a typical \bb candidate selection. A second sample, with statistics corresponding to an exposure of 10 year of data-taking, has been generated without simulating the \Po{214} decay. This sample is used to derive a probability density function (PDF) of the energy spectrum of the \Bi{214} background with fine details. The reconstruction follows the lines described in Sec.~\ref{sec:alphareco}, but avoids the assumption of point-like events, as electron tracks have lengths in the order of tens of cm. In this case, the SiPM signals provide the X and Y coordinates of the event for each time (or Z) slice of the S2 signals. The event energy reconstruction considers the corrections associated to the electron lifetime and the light collection variation across the XY plane described, using a simulation of \Kr events in the active volume. The electron-like events are selected by requiring only one S1, with a yield not exceeding 900 p.e. In addition, the events are required to be fully contained within the active volume, by imposing only one S2 with all associated hits inside being at least 40 mm away from the TPC boundaries (Z$\in$[40,1147]~mm and R<452 mm). Hereafter, the resulting sample is referred to as the fiducial electron selection. Figure~\ref{fig:bi214} shows the energy spectra of the inclusive and fiducial selection samples from the MC simulation of \Bi{214} on the cathode. The corresponding rates derived from the MC sample accounting for the \Bi{214}$\rightarrow$\Po{214} chain have been used to normalize the \Bi{214} PDF. It is worth noticing that all the remaining events in the fiducial sample are gamma-induced, as 100\% of the $\beta$ and $\alpha$ particles emitted from the cathode surface and entering the TPC volume produce hits outside the fiducial region.

The fiducial selection leads to a rate of R$^{\text{fid}}$(\Bi{214})=(0.26$\pm$0.03(stat)$\pm$0.03(sys))~mHz for energies above 500 keV. The systematic uncertainty accounts for the error associated with the estimation of the rate of \Bi{214} decays at the cathode ($\sim$11\%) presented in Sec.~\ref{sec:bicath}. This value is used to compute the background index in a region of interest (ROI) for the \bbnonu search, defined between 2.4 MeV and 2.5 MeV. The fiducial background index in ROI yields a value of (7.3$\pm$1.5(stat)$\pm$0.8(sys))$\times10^{-4}$ counts/(keV$\cdot$kg$\cdot$yr), associated with a rejection factor of (99.979$\pm$0.005)\%. As described in \cite{NEXT:2021dqj,NEXT:2023daz}, the selection of the \bb candidates in the NEXT detectors considers the fiducialization of the events, as well as a selection based on the reconstructed tracks. This topological selection consists of requiring only one track in the event, with the track being a double-electron candidate (see Ref.~\cite{NEXT:2020try}). Although the detailed study and optimization of the topological-based selection are beyond the scope of this work, reasonable approaches are considered to estimate the eventual background index of the Rn-induced \Bi{214}. In order to select events with one track, those ones with more than one cluster of SiPM hits are rejected. This yields an extra rejection factor of (43.5$\pm$10.3)\% for events within the ROI. Furthermore, in consideration of the results presented in Ref.~\cite{NEXT:2020try}, a conservative rejection factor of 90\% for single-electron candidates is assumed. Overall, the background index due to the Rn-induced \Bi{214} in the low-pressure run is $\sim$4$\times10^{-5}$ counts/(keV$\cdot$kg$\cdot$yr). Given the expected total radiogenic background index in NEXT-100, 4$\times10^{-4}$ counts/(keV$\cdot$kg$\cdot$yr) \cite{NEXT:2015wlq}, it is concluded that the internal radon will not induce a significant background in the search for the \bbnonu decay. A summary of the rejection factors and background indices for each step in the \bb-like selection is presented in Tab.~\ref{tab:bgindex}.   

\begin{table}[!htb]
\caption{\label{tab:bgindex} Rejection factors and cumulative background indices for Rn-induced \Bi{214}.}
\begin{center} 
\begin{tabular}{c|cc}
\hline
Selection & Rejection factor (\%) & Background index  ($10^{-4}$ c/(keV$\cdot$kg$\cdot$yr))\\ \hline
Fiducial  & 99.979$\pm$0.005  & 7.3$\pm$1.5(stat)$\pm$0.8(sys)  \\
1 track   & 43.5$\pm$10.3     & 4.2$\pm$1.1(stat)$\pm$0.5(sys)  \\
double-e$^-$ track & $\sim$90 & $\sim$0.4                        \\
\hline
\end{tabular}
\end{center}
\end{table}

\section{Airborne Rn-induced background}
\label{sec:airborne}
The radon concentration in the air of the HALL A of the LSC is continuously monitored with an AlphaGuard detector (AlphaGuard Saphymo P30). This detector has a sensitivity of 1 cpm at 20 Bq/m$^3$, a lower limit of $\sim$2 Bq/m$^3$, and a calibration error of 3\%. The average annual activity is (69.0$\pm$0.3)~Bq/m$^3$, with daily, seasonal and annual ﬂuctuations typically ranging from $\sim$60 Bq/m$^3$ to $\sim$110 Bq/m$^3$ \cite{Bandac:2017eks,Perez-Perez:2021bzu}. These variations have been exploited in NEXT-100 in order to evaluate the impact of the airborne radon in the background budget of the experiment. In particular, the correlation between the radon concentration in the air and the rate of electron-like fiducial events collected by NEXT-100 has been analyzed during two low-background data-taking periods, with and without the RAS in operation (E2 and E1, respectively).

\subsection{Reconstruction and selection of electron events}
\label{sec:elecreco}

The reconstruction follows the same procedure applied to the \Bi{214} MC in Sec.~\ref{sec:bb0nubg}. After reconstructing the S1 and S2 signals in the global PMT waveform, 3D hits are reconstructed with the SiPM hits found in each time slice of the S2 signals. According to the amplitude of the signal in each SiPM, the energy measured by the PMTs is distributed among all the reconstructed hits. The energy of each hit is then corrected for the light collection variations in the XY plane and the electron attachment along the drift axis (Z), using the \Kr data. These corrections have been described in detail in \cite{NEXT:2025fpq,NEXT:2025ozn}. The total energy is computed as the sum of the energy of all the hits in the event. To account for the non-linearities in the energy response, the energy scale for extended objects is obtained from the analysis of \Th{228} calibration data, taken under the same detector operation conditions as in the low-background period. The \Th{228} source, deployed in a dedicated port on the pressure vessel, allows for the reconstruction of an extended energy spectrum exhibiting the characteristic gamma lines of \Tl{208} (510.7 keV, 583 keV, 860 keV and 2615 keV), as well as the double escape peak (1593 keV), and the \Bi{212} 727 keV $\gamma$ line. These are fitted to an empirical linear model in order to determine the energy scale as a function of the energy. For the analysis of the radon induced background, hereafter we only consider events with a calibrated energy above 0.5 MeV. 

\begin{figure}
  \begin{center}
    \includegraphics[scale=0.30]{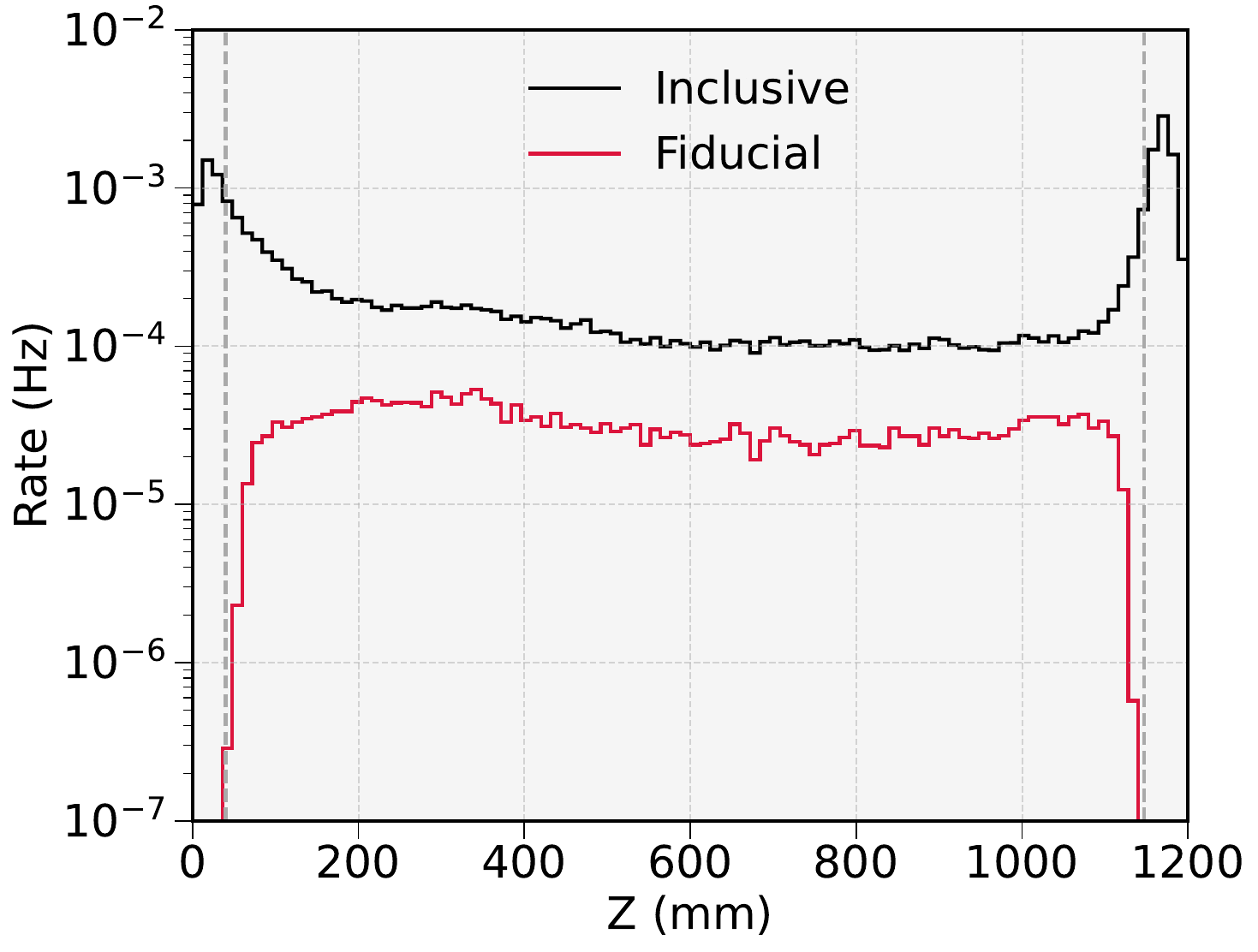}
    \includegraphics[scale=0.30]{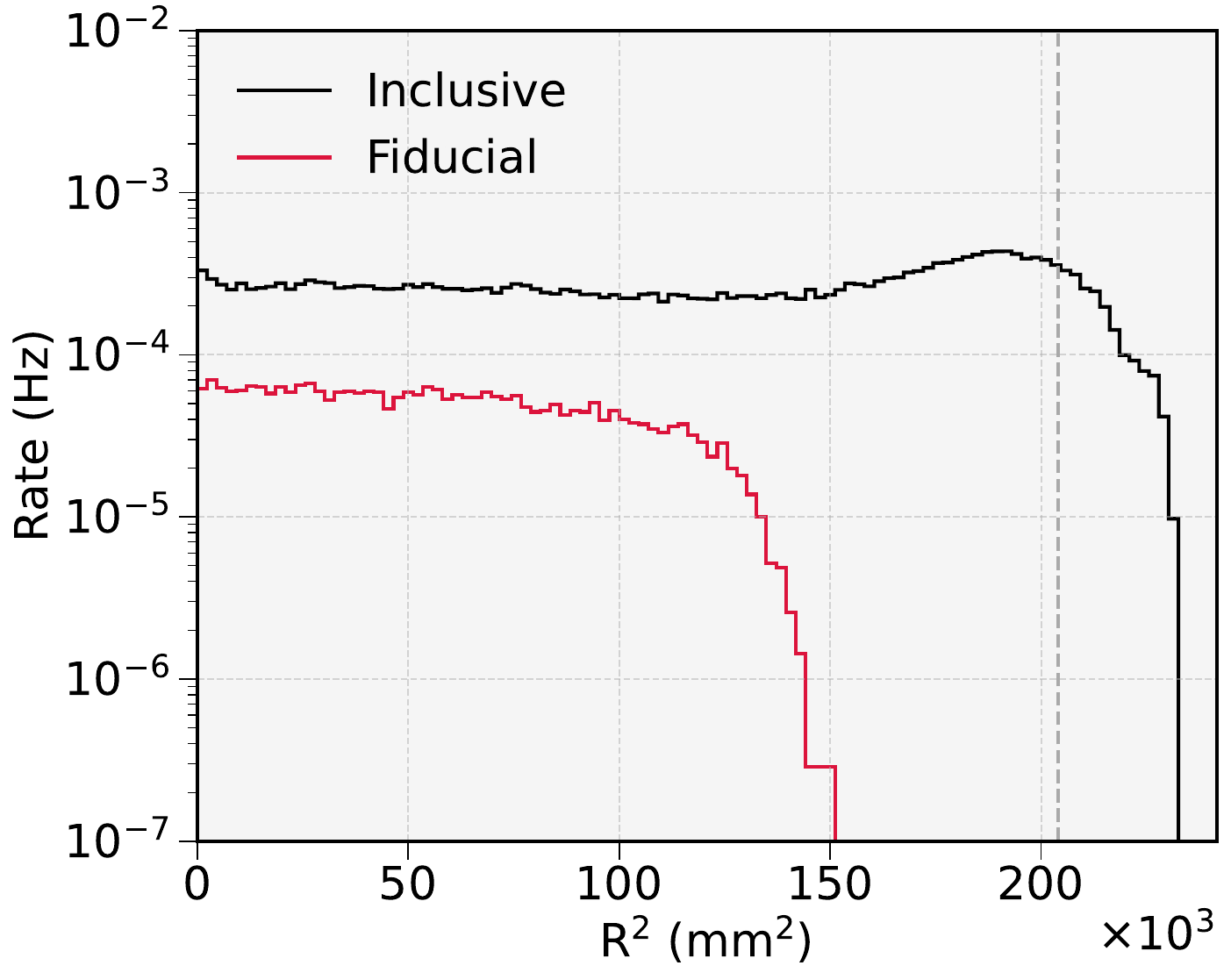}
    \caption{Charge-weighted average Z (left) and R$^2$ (right) distributions of the electron-like events in the low-background data (E2). The solid red (black) line corresponds to fiducial (inclusive) sample. The dashed vertical lines mark the fiducial cuts applied to the individual hits in the events.}
    \label{fig:edist}
  \end{center}
\end{figure}

From the reconstructed data, alpha decay events are rejected by a cut on the maximum energy of the S1 signal (900 p.e.). As done for the \Bi{214} MC in Sec.\ref{sec:bb0nubg}, the remaining electron-like events undergo a fiducial selection requiring only one S2 and all hits contained within the volume defined by Z$\in$[40,1147] mm and R<452 mm. The charge-weighted average positions of the inclusive and selected fiducial electron-like events during E2 are shown in Fig.~\ref{fig:edist}. The background reduction of the fiducial selection, which removes the electron background emanating from the surfaces of the active volume, is clearly visible. In particular, the peak of events at the cathode position, dominated by Rn-induced \Bi{214}, is significantly suppressed.

\subsection{Correlation with radon activity in the LSC}
\label{sec:rncorr}


The average fiducial rate of events during the E1 period is (3.81$\pm$0.07)~mHz. Due to the variations in the radon concentration in the air of the LSC, this fiducial rate exhibits fluctuations of $\sim$20\%. During the E2 period, a constant fiducial rate of (2.90$\pm$0.04)~mHz has been measured. Thus, the contribution of the airborne-Rn-induced background when the RAS is not in operation yields an average of (0.91$\pm$0.08)~mHz. This contribution corresponds to the $\gamma$s involved in the \Bi{214} decays, which reach the active volume through the pressure vessel and the inner copper shield of NEXT-100.  

\begin{figure}
  \begin{center}
    \includegraphics[scale=0.30]{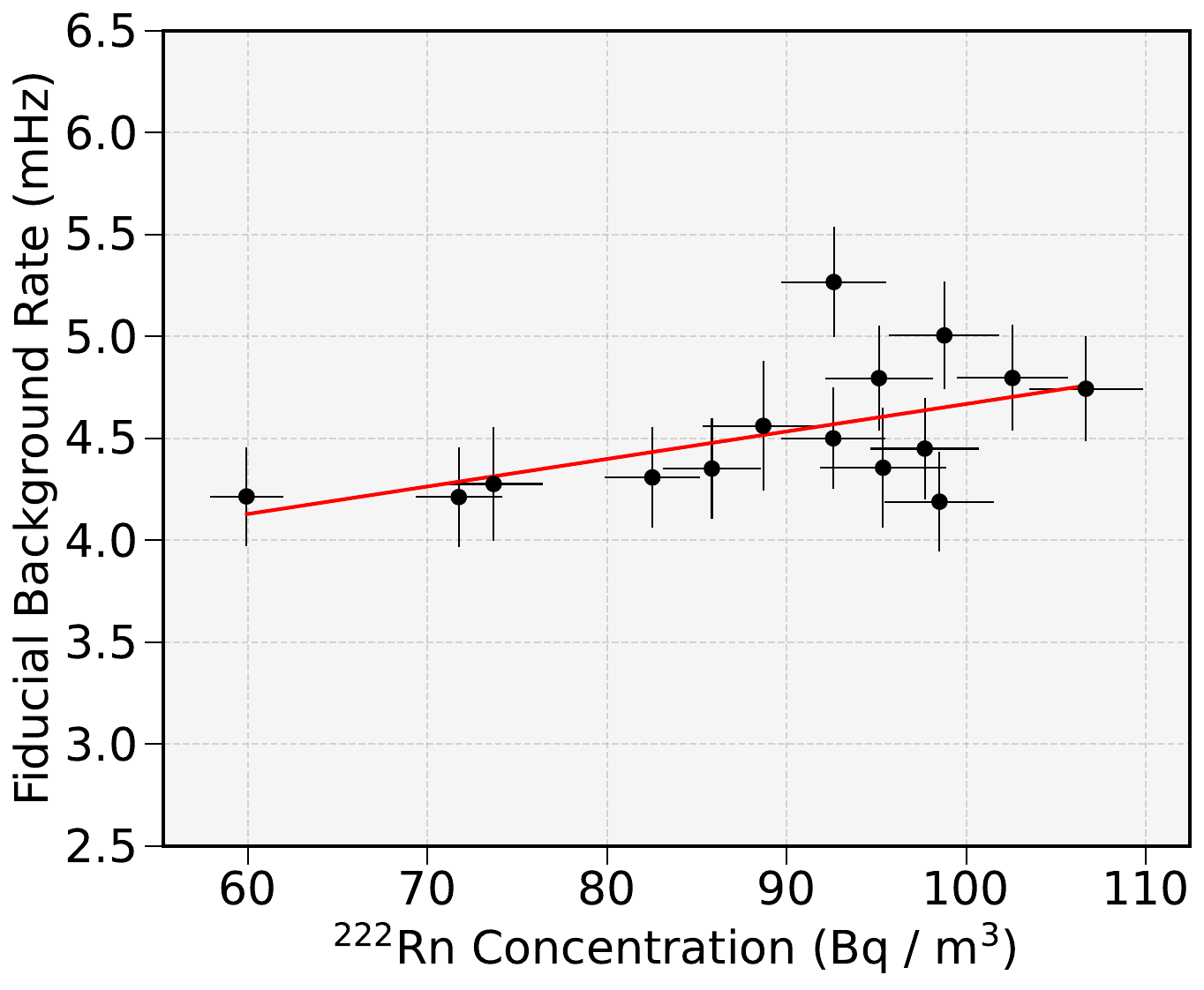}
    \includegraphics[scale=0.30]{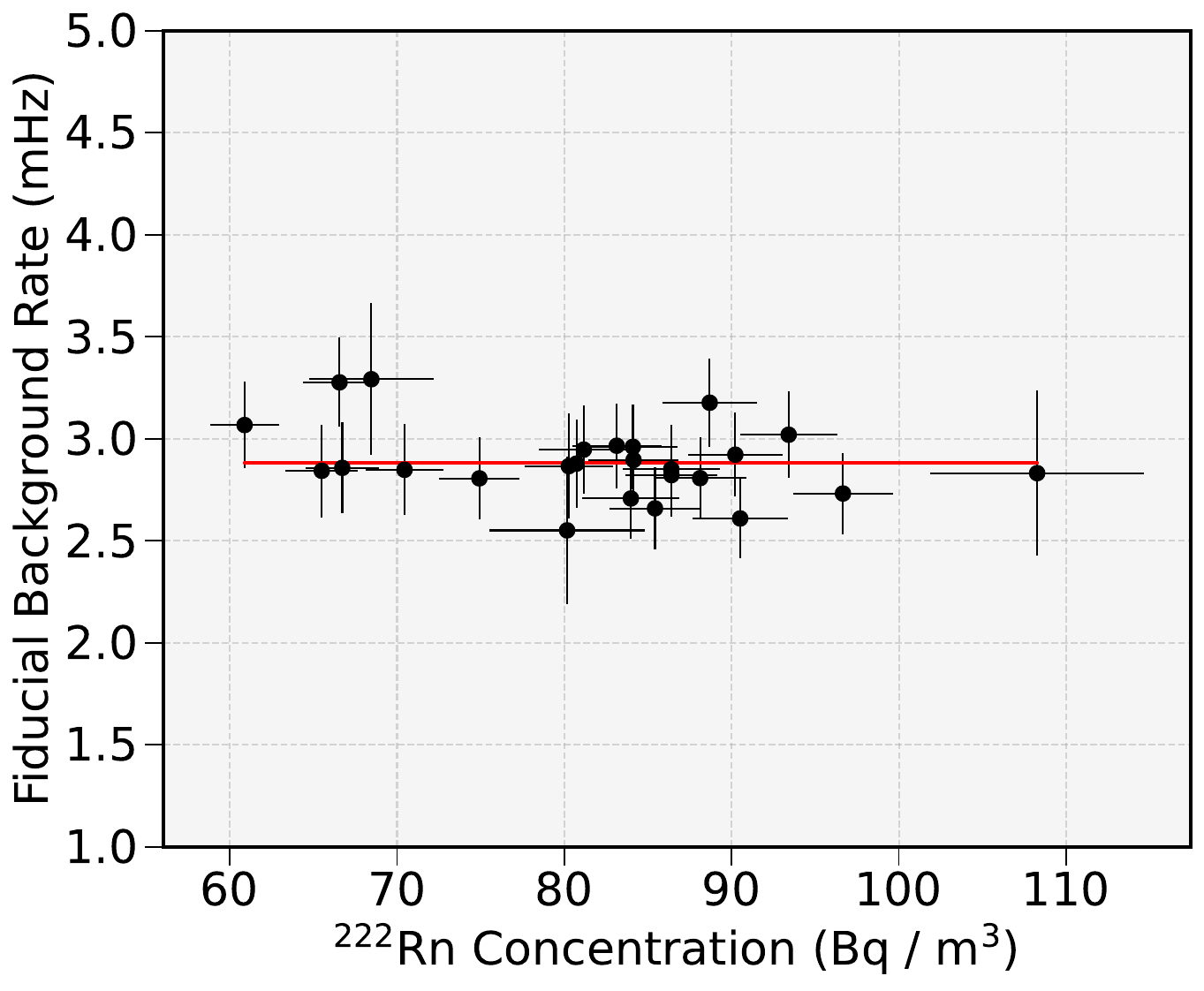}
    \caption{Correlation between the rate of the fiducial events in NEXT-100 and the radon activity in the HALL A of the LSC. Left: correlation without the RAS in operation (E1), fitted to a linear function  which provides an estimate of the rate of fiducial events in absence of airborne \Rn{222}. Right: correlation with the RAS in operation (E2), fitted to a constant rate.}
    \label{fig:rncorr}
  \end{center}
\end{figure}

During the E1 and E2 data-taking periods, runs of about 24 hours have been collected. The radon activity in the HALL A of the LSC during each one of these runs has been computed as the average of the AlphaGuard measurements taken every 10 minutes. The correlation between the rate of fiducial events in NEXT-100 and the radon activity in the surrounding air is presented in Fig.~\ref{fig:rncorr} for the periods with and without the RAS in operation. During E1 (left panel of Fig.~\ref{fig:rncorr}), the increasing trend of the rate with the radon activity is fitted to a linear function. From the best-fit intercept, the expected fiducial background rate in absence of radon activity is estimated: (3.32$\pm$0.47)~mHz. The data collected with the RAS in operation exhibit a significant reduction in the fiducial background rate, which has been found to be stable in time. In particular, no significant correlation with the radon activity in the LSC is observed (right panel of Fig.~\ref{fig:rncorr}). A fit to a constant value yields a rate of (2.88$\pm$0.04)~mHz, with a $\chi^2/\text{ndof}$=0.6 (p-value=95\%). This measurement of the fiducial background in NEXT-100 in E2 is fully consistent with the previous expectation in absence of radon activity, derived from the E1 data. Overall, NEXT-100 is not sensitive to the impact of the airborne \Rn{222} in the LSC when the RAS is flushing radon-free air inside the lead castle. Thus, this radon background contribution is not expected to be significant in the search for the \bbnonu.      

\section{Summary and conclusions}
\label{sec:conc}

The NEXT-100 detector aims at performing the first competitive search for the \bbnonu decay using a high-pressure electroluminescent \Xe{136} TPC. After being installed and commissioned at the LSC during 2024, the detector has taken a first physics run in 2025 with \Xe{136}-depleted xenon at 3.95 bar. The first low-background data samples have been devoted to the measurement and characterization of the radiogenic backgrounds, and in particular to the evaluation of the radon-related ones. In this work, the impact of the Rn-induced background in the \bbnonu search has been evaluated with data taken under different operation conditions. This study has considered both internal radon, from detector material emanation, and airborne radon, present in the air of the HALL A of the LSC.

In order to address the internal radon contribution, dedicated alpha runs have been taken in two different periods: a high radon activity run recirculating the xenon gas through a cold getter in the gas system (from which large amounts of $^{222}$Rn emanate), and a low radon activity run, purifying the gas through a hot getter. The decrease of the trigger rate as a function of time during the second period has allowed for the determination of a half-life of (3.77$\pm$0.06)~day, fully consistent with the one of $^{222}$Rn. The intrinsic internal radon contamination when using the hot getter has been measured to be (0.95$\pm$0.04(stat)$\pm$0.09(sys))~Bq/m$^3$. From the analysis of the energy spectrum of the alpha decays, it is derived that $\sim$93\% of the Rn-induced \Bi{214} plate-out on the cathode surface, resulting in an activity of (0.97$\pm$0.05(stat)$\pm$0.10(sys))~Hz. By means of a MC simulation, this is estimated to imply a fiducial background rate of (0.26$\pm$0.03(stat)$\pm${0.03}(sys))~mHz. For a \bbnonu ROI between 2.4 MeV and 2.5 MeV, this corresponds to a background index of (7.3$\pm$1.5(stat)$\pm$0.8(sys))$\times10^{-4}$~counts/(keV$\cdot$kg$\cdot$yr). After a topological selection requiring only a double-electron-like track in the events, this value is reduced down to $\sim$4$\times10^{-5}$~counts/(keV$\cdot$kg$\cdot$yr). As the expected background budget in NEXT-100 is\linebreak 4$\times10^{-4}$counts/(keV$\cdot$kg$\cdot$yr), this analysis demonstrates that the internal radon will not be a significant background in the \bbnonu search.

It is worth noticing that the measured activity of \Rn{222} in NEXT-100 is significantly higher than the value reported for the NEXT-White detector of (38.1$\pm$6.3)~mBq/m$^3$~\cite{Novella:2018ewv}. The origin of this difference is not understood yet and it is currently under investigation. Most of the internal radon comes from the gas system materials, as they are not radiopure. However, the gas system is essentially the same as in the operation of the NEXT-White detector. The level of internal radon will be measured again when operating NEXT-100 at 10 bar (nominal pressure for the \bbnonu search). This will allow for a comparison with the NEXT-White results to be performed at the same operating conditions. Should the origin of this excess be identified and suppressed, the Rn-induced background index measured in this work might be significantly reduced.   

The background induced by the airborne radon, within the detector and the lead castle walls, has been evaluated relying on two low-background runs with the operation and trigger conditions of a physics run. While the radon abatement system of the LSC was not in operation during the first run, it was delivering radon-free air during the second. From the correlation between the fiducial background rate and the measurements of the the radon activity in HALL A, a background rate of (3.32$\pm$0.47)~mHz in absence of airborne radon has been estimated from the RAS-Off run. This rate is fully consistent with the observed fiducial rate in the RAS-On run: (2.88$\pm$0.04)~mHz. In addition, this value has been stable with time, showing no correlation with the radon activity in HALL A. Overall, it is concluded that the NEXT-100 detector operates in a virtually radon-free air when the RAS is delivering air inside the lead castle. This is consistent with the study performed with NEXT-White~\cite{Novella:2019cne}, as it might be expected as the lead castle and radon abatement system are the same for both detectors.


\acknowledgments
Authors acknowledge the data as well as the technical advice provided in the framework of the collaboration between the Canfranc Underground Laboratory and the LABAC (University of Zaragoza). The NEXT Collaboration acknowledges support from the following agencies and institutions: the European Research Council (ERC) under Grant Agreement No. 951281-BOLD and 101039048-GanESS; the European Union’s Framework Programme for Research and Innovation Horizon 2020 (2014–2020) under Grant Agreement No. 860881-HIDDeN; the MCIN/AEI of Spain and ERDF A way of making Europe under grants PID2021-125475NB and RTI2018-095979, and the Severo Ochoa and Mar\'ia de Maeztu Program grants CEX2023-001292-S, CEX2023-001318-M and CEX2018-000867-S; the Generalitat Valenciana of Spain under grants PROMETEO/2021/087, ASFAE/2022/028, ASFAE/2022/029, CISEJI/2023/27 and CIDEXG/2023/16; the Department of Education of the Basque Government of Spain under the predoctoral training program non-doctoral research personnel; the Spanish la Caixa Foundation (ID 100010434) under fellowship code LCF/BQ/PI22/11910019; the Portuguese FCT under project UID/FIS/04559/2020 to fund the activities of LIBPhys-UC; the Israel Science Foundation (ISF) under grant 1223/21; the Pazy Foundation (Israel) under grants 310/22, 315/19 and 465; the US Department of Energy under contracts number DE-AC02-06CH11357 (Argonne National Laboratory), DE-FG02-13ER42020 (Texas A\&M), DE-SC0019054 (Texas Arlington) and DE-SC0019223 (Texas Arlington); the US National Science Foundation under award number NSF CHE 2004111; the Robert A Welch Foundation under award number Y-2031-20200401. C. Romo-Luque acknowledges financial support from LANL to participate in NEXT operations. Finally, we are grateful to the Laboratorio Subterr\'aneo de Canfranc for hosting and supporting the NEXT experiment.




\bibliographystyle{JHEP}
\bibliography{biblio}

\end{document}